\documentclass[pre,preprint,numerical,nofootinbib]{revtex4-1}

\usepackage{amsmath}
\usepackage{graphicx}
\usepackage{amsfonts}
\usepackage[usenames,dvipsnames]{color}
\usepackage{bm} 
\usepackage{todonotes}
\usepackage{hhline} 
\usepackage{hyperref}
\usepackage{float}

\usepackage{algorithm} 
\usepackage[noend]{algpseudocode}

\newcommand{\vect}[1]{\mbox{\boldmath $#1$}}

\begin{document}

\title{Topological data analysis of continuum percolation with disks}

\author{Leo Speidel}
\affiliation{Department of Statistics, University of Oxford, Oxford, UK}
\affiliation{Doctoral Training center in Systems Biology, University of Oxford, Oxford, UK}

\author{Heather A. Harrington}
\affiliation{Mathematical Institute, University of Oxford, Oxford, UK}

\author{S. Jonathan Chapman}
\affiliation{Mathematical Institute, University of Oxford, Oxford, UK}

\author{Mason A. Porter}
\affiliation{Department of Mathematics, University of California Los Angeles, Los Angeles, CA, USA}

\date{\today}

\begin{abstract}

We study continuum percolation with disks, a variant of continuum percolation in two-dimensional Euclidean space, by applying tools from topological data analysis. We interpret each realization of continuum percolation with disks as a topological subspace of $[0,1]^2$ and investigate its topological features across many realizations. We apply persistent homology to investigate topological changes as we vary the number and radius of disks. We observe evidence that the longest persisting invariant is born at or near the percolation transition.
\end{abstract}

\maketitle



\section{Introduction}

Percolation, and its many variants, is one of the most popular topics in statistical physics. It has been used in the study of a diverse variety of phenomena, ranging from biological and social contagions to connectivity of infrastructure \cite{saberi2015,frontiers2016,kesten-whatis}. Continuum percolation models can help provide a mechanistic understanding for the formation of structures embedded in space and are closely related to spatially-embedded random graphs~\cite{Penrose2003}. Understanding the properties of spatially-embedded random graphs has long been of theoretical interest~\cite{Meester1996,Gilbert1961,Penrose2003,Kahle2014,Kahle2011}, and it also has numerous applications, such as for investigating properties of granular packings \cite{bassett2012influence,Setford2014} and sensor networks~\cite{Nemeth2003,Dettmann2016,dettmann2018}. 

Studying spatial percolation processes can be helpful for investigations of spatially embedded systems, which is particularly important in light of the increased availability of rich, finely-grained data about such systems. For example, advances in imaging technology have led to increasingly accurate measurements of biophysical networks~\cite{Kherlopian2008} (such as vascular networks, leaf-venation networks, bronchial trees in the lung, and neuronal networks), which --- like many other networks and complex systems --- are embedded in space \cite{Barthelemy2011}. Their properties and mechanisms of the formation of biophysical networks have been investigated as
optimization problems~\cite{Katefori2010}, and they have also been studied using fluid dynamics~\cite{Pries1996} and in relation to random processes such as random sequential adsorption (RSA)~\cite{Evans1993}. Some of these situations, such as flocks of \emph{Volvox barberi} \cite{balas2018}, also form fascinating packing structures.

In many continuum percolation models (and in percolation problems more generally \cite{saberi2015,frontiers2016}), there are known phase transitions (identified theoretically or computationally) as one varies the model parameters~\cite{Meester1996,Penrose2003}. For instance, a large percolating cluster may suddenly emerge as smaller clusters connect to each other. Percolation transitions are interesting, because they indicate a change in long-range spatial correlations. To detect such structural changes, one commonly studies changes in the sizes of connected clusters. 

In the present paper, we take a different approach, as we focus instead on the \emph{shape} of clusters. We do computations to investigate changes in the topological\footnote{We use the word ``topology'' in its standard mathematical sense, rather than in the sense of examining connectivity structure in networks (``graph topology'' or ``network topology'') that is often used to describe the structure of graphs and networks.} properties of clusters~\cite{Edelsbrunner2010, Kaczynski2006computational, Otter2015}. The extent to which varying model parameters affects topological properties has been examined in some classical models of statistical physics, such as in models of interacting spins~\cite{Pettini2007geometry, Donato2016PHXY, Santos2017topological}, and we do this in our focal problem as well. Specifically, we study one-dimensional topological invariants, which one can interpret as cycles or holes in the structure of a point cloud of data (including data from a network). We examine what we call a ``continuum percolation with disks'', a variant of continuum percolation, in which we place $N$ disks of radius $r$ independently and uniformly at random in two-dimensional (2D) Euclidian space. This percolation model is related to random geometric graphs (RGGs)~\cite{Gilbert1961, DallPRE2002, Penrose2003, Dettmann2016}.

To identify topological invariants of continuum percolation with disks, we calculate its homology groups $H_0$ and $H_1$. The dimension of $H_0$ equals the number of connected components, and we note that the percolation and connectivity transitions are phase transitions related to $H_0$. The percolation transition occurs in the regime $0 < \lim_{N \to \infty }Nr^2 < \infty$, and the connectivity transition occurs in the regime $\lim_{N \to \infty }Nr^2 = \infty$~\cite{Penrose2003}. The latter regime is thus in the supercritical regime of the percolation transition. One can interpret the dimension of $H_1$ as the number of cycles, in the form of one-dimensional holes, in a structure. Previous studies have identified two topological phase transitions that are related to the first homology group $H_1$. They each correspond to a transition between zero and nonzero dimension of $H_1$ \cite{Kahle2011, Kahle2014}. The first phase transition occurs in the subcritical regime of the percolation transition (in which $\lim_{N \to \infty }Nr^2 = 0$), and the second phase transition occurs in the supercritical regime of the percolation transition.

It is partially understood how the first homology group changes in the critical regime of percolation transition (see Section~\ref{sec:known_results}), but it is not well-understood how it changes at and near the percolation transition. We apply persistent homology (PH) \cite{Edelsbrunner2010,Otter2015} and study topological changes close to the percolation transition. Based on our calculations, the longest persisting topological invariant appears to be born at the percolation transition. This suggests that there is a change in shape, in addition to a change in size, of clusters at the percolation transition.

The rest of our paper is organized as follows. First, we define continuum percolation more precisely and then define \emph{continuum percolation with disks}. To help characterize continuum percolation with disks, we then discuss the sizes, number, and shapes of its clusters. We investigate continuum percolation with disks by computing PH and then conclude with a brief summary and discussion.


\section{Continuum percolation}

\begin{figure*}
  \centering
  \includegraphics[width=\textwidth]{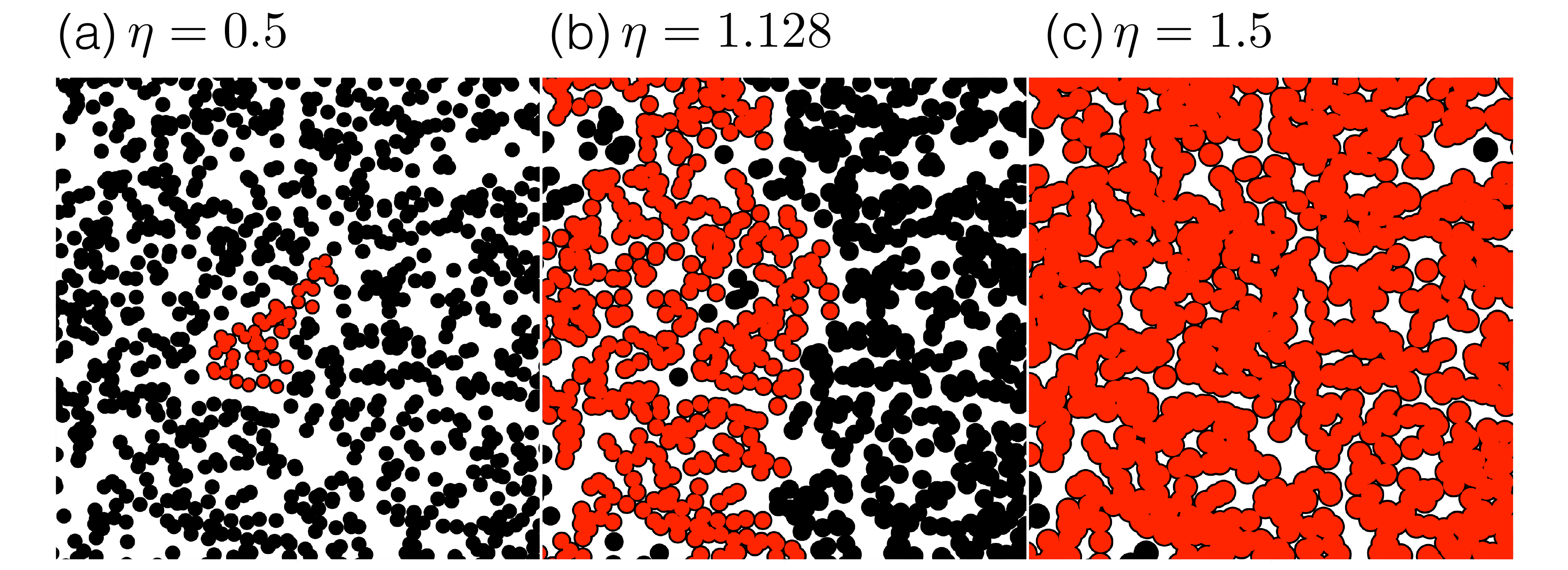}
  \caption{
    \label{fig:discs_N1000}
       One realization of continuum percolation with disks with filling factors of (a) $\eta(N,r) = 0.5$, (b) $\eta(N,r) = 1.128$, and (c) $\eta(N,r) = 1.5$, where $\eta(N,r) = N \pi r^2$ and there are $N=1000$ disks of radius $r$. We show the largest connected cluster in red. It has been estimated that a percolation transition occurs at $\eta_{\rm c} \approx 1.128$~\cite{Mertens2012}.
  }
\end{figure*}

Consider a metric space $(X,d)$. \emph{Continuum-percolation} models are random processes in which subsets $D_i \subseteq X$ (where $i \in \{1,\ldots,N\}$) are chosen randomly with some probability distribution, resulting in a structure $\bigcup_{i = 1}^{N} D_i$.

We consider the unit square $X=[0,1]^2$ equipped with the Euclidean metric. We sample points $\vect{x}_i \in X$ (with $i \in \{1, \ldots, N\}$) independently and uniformly at random and obtain disks $D_i(r) = \{\vect{y} \in X: d(\vect{y}, \vect{x}_i) < r\}$ of radius $r \ge 0$ that are centered at these points. The radius $r$ is the same for all disks. The resulting structure $\mathcal{D}(r) = \bigcup_{i = 1}^{N} D_i(r)$ is a union of potentially overlapping disks (see Fig.~\ref{fig:discs_N1000}). We refer to this model as \emph{continuum percolation with disks}. 

A related model is a random geometric graph (RGG)~\cite{Dettmann2016, Penrose2016}. One can construct an RGG from a continuum-percolation model by defining the node set $V = \{1, \ldots, N\}$ and edge set $E = \{ (i,j) : D_i \cap D_j \neq \emptyset \}$. One can induce an RGG from continuum percolation by placing nodes independently and uniformly at random in 2D Euclidean space and connecting pairs of nodes that lie within a specified distance $2r$~\cite{Gilbert1961, DallPRE2002}. 

Both continuum percolation and RGGs have been studied in a variety of metric spaces~\cite{DallPRE2002,Barthelemy2011, Rintoul1997, Penrose2003}. One can also choose different probability distributions for placing points; consider higher-dimensional spaces; and consider ``softer'' potentials (in contrast to the Heaviside function above), in which nodes are connected to each other with a probability that decays with increasing distance between them \cite{Dettmann2016}.


\section{Sizes and number of clusters}
\label{sec:size}

One can partition any realization of the structure $\mathcal{D}(r)$ into disjoint clusters. The distributions of the number of clusters and the number of disks in the largest cluster depend on the combination of the parameters $N$ and $r$.

It is useful to study $\eta(N,r) = N \pi r^2$, which is called the \emph{filling factor}. The size of the largest cluster undergoes a phase transition as one varies $\eta(N,r)$. For $\lim_{N \to \infty} \eta(N,r) < \eta_{\rm c}$, the size of the largest cluster is, with probability $1$, at most~$O(\log N)$ of the number $N$ of disks as $N \to \infty$ (see Theorem 10.3 in Ref.~\cite{Penrose2003}). When $\lim_{N \to \infty}\eta(N,r)$ approaches the critical percolation transition value (i.e., the percolation ``threshold'') $\eta_{\rm c}$ from below, clusters merge until, at $\lim_{N \to \infty} \eta(N,r) = \eta_{\rm c}$, there emerges a unique giant cluster (i.e., giant component) with size~$O(N)$~(see Theorem 10.9 in Ref.~\cite{Penrose2003}). Percolation transitions have been studied for many models and in numerous situations, including for lattice models and networks, continuum models with various geometric objects, and in different dimensions~\cite{Mertens2012, Grimmett1999, Bollobas2006,frontiers2016,Newman2010}. For continuum percolation with disks, the exact value $\eta_{\rm c}$ of the percolation threshold is unknown. It has been estimated numerically to be $\eta_{\rm c} \approx 1.128$~\cite{Mertens2012}.

In the supercritical regime $\lim_{N \to \infty}\eta(N,r) > \eta_{\rm c}$, clusters keep merging as $\eta(N,r)$ increases. In a second (``connectivity'') phase transition, which occurs in the limit $\lim_{N \to \infty}\eta(N,r) = \infty$, all clusters merge into a single cluster with probability $1$. The critical value for this connectivity transition satisfies~\cite{Penrose2003}
\begin{equation} \label{eq:connectivity_transition}
 	 \lim_{N \to \infty} \frac{\eta_{\rm c}^{\rm connect}(N,r)}{\log N} = \frac{1}{4}\,.
\end{equation}


\section{Shape of clusters}
\label{sec:shape}

\begin{figure*}
  \centering
  \includegraphics[width=\textwidth]{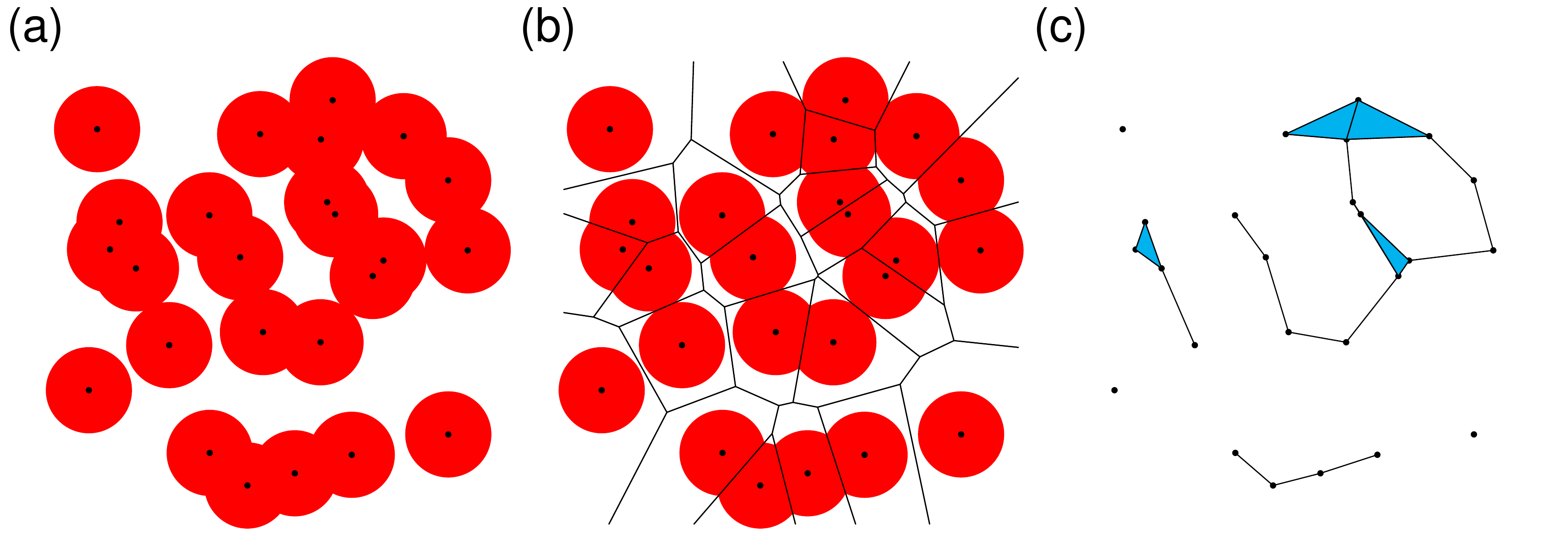}
  \caption{
    \label{fig:alpha_complex}
    (a) A realization of continuum percolation with disks with $N=25$ and $r = 0.22$. (b) Voronoi tessellation constructed using the center of the disks in panel (a). (c) Alpha complex constructed from panel (b). We represent 0-simplices by nodes, 1-simplices by edges (which connect adjacent nodes to each other), and 2-simplicies by filled triangles. In this example, there are six clusters (so $\beta_0 = 6$) and one hole (so $\beta_1 = 1$).
  }
\end{figure*}

We are interested in understanding the change in shape of connected clusters as we vary $\eta(N,r)$. To investigate this, we interpret $\mathcal{D}(r)$ as a topological subspace of $(X,d)$ and study its topological properties. Specifically, we study its \emph{homology groups}, which correspond, intuitively, to the numbers of holes in different dimensions~\cite{Hatcher2002algebraic}.

We can calculate the homology groups of $\mathcal{D}(r)$ by studying a different topological space with homology groups that are isomorphic to those of $\mathcal{D}(r)$. The topological space that we study is the \emph{alpha complex}~\cite{Edelsbrunner1983}, which is a type of \emph{simplicial complex}. (See e.g., Refs.~\cite{Hatcher2002algebraic,Otter2015} for the definition of a simplicial complex.) To construct an alpha complex, we calculate the Voronoi tesselation from the centers of the disks. We define the intersection of a disk and its corresponding Voronoi cell by $V_i(r)$ (where $i \in \{1,\ldots,N\}$). The alpha complex includes 0-simplices $\{V_i(r)\}$ (where $i \in \{1,\ldots,N\}$) and $k$-simplices (with $k = 1$ and $k=2$), given by $\{V_{i_0}(r), \ldots, V_{i_k}(r)\}$, such that $V_{i_0}(r) \cap \cdots \cap V_{i_k}(r) \neq \emptyset$. In our setting, the Nerve Theorem implies that $\mathcal{C}(r)$ and $\mathcal{D}(r)$ have the same homology (i.e., their homology groups are isomorphic)~\cite{Edelsbrunner2010, Bjorner1995Chapter}.  In Fig.~\ref{fig:betti_number}, we visualize the alpha complex for a small example of continuum percolation with disks. 

The $k$th homology group $H_k(\mathcal{C}(r))$ describes the $k$-dimensional holes~\cite{Hatcher2002algebraic,Otter2015}. Its dimension is the $k$th Betti number $\beta_k$. The $0$th Betti number $\beta_0$ is equal to the number of clusters. As we discussed in Section~\ref{sec:size}, $\beta_0$ has been studied extensively in percolation (though without explicitly invoking ideas from homology). Intuitively, the first Betti number $\beta_1$ counts the number of holes. In this paper, we focus on $\beta_1$, which gives information about the ``shapes'' of the clusters.


\subsection{Known phase transitions in the first Betti number}
\label{sec:known_results}

Theoretical work relating Betti numbers to continuum percolation indicates that the shapes of clusters change as one varies $\eta(N,r)$ ~\cite{Kahle2014, Kahle2011}. For instance, it has been shown, in the limit $\lim_{N \to \infty} \eta(N,r) = 0$, that there is a phase transition at which nontrivial 1D homology first appears. The critical value satisfies
\begin{equation}
	  \lim_{N \to \infty} \sqrt{N}\eta_{\rm c}^{\rm exist}(N,r) = c_{\rm exist}\,.
\end{equation}
Below the critical value, $\beta_1 = 0$; above the critical value, $\beta_1 > 0$ with probability $1$ (see, e.g., Theorem 4.12~in Ref.~\cite{Kahle2014}). 
 
Similar to the connectivity transition in Eq.~\eqref{eq:connectivity_transition}, the first Betti number vanishes in the limit $\lim_{N \to \infty} \eta(N,r) = \infty$. The critical value for this phase transition satisfies (see, e.g., Theorem 4.12~in Ref.~\cite{Kahle2014})
\begin{equation}
	  \lim_{N \to \infty} \frac{\eta_{\rm c}^{\rm vanish}(N,r)}{\log N} = c_{\rm vanish}\,.
\end{equation}
In the critical regime, in which $\eta(N,r)$ approaches a finite, nonzero limit as $N \to \infty$, it is known that $\beta_1$ scales linearly in the number of disks:
\begin{equation}  \label{eq:linear_scaling_beta1}
  	\lim_{N \to \infty} \frac{E[\beta_1]}{N} = f(\eta)\,,
\end{equation}
for some function $f(\eta) \in (0,\infty)$. (See, e.g., Theorem 4.6~in Ref.~\cite{Kahle2014}.) Other than Eq.~\eqref{eq:linear_scaling_beta1}, it is not understood what happens at or near the percolation transition. In Section \ref{sec:TDA}, we confirm Eq.~\eqref{eq:linear_scaling_beta1} (see Fig.~\ref{fig:betti_number}) and investigate the behavior of $f(\eta)$ near the percolation transition.


\section{Persistent homology and results}
\label{sec:TDA}

\begin{figure*}
  \centering
  \includegraphics[width=\textwidth]{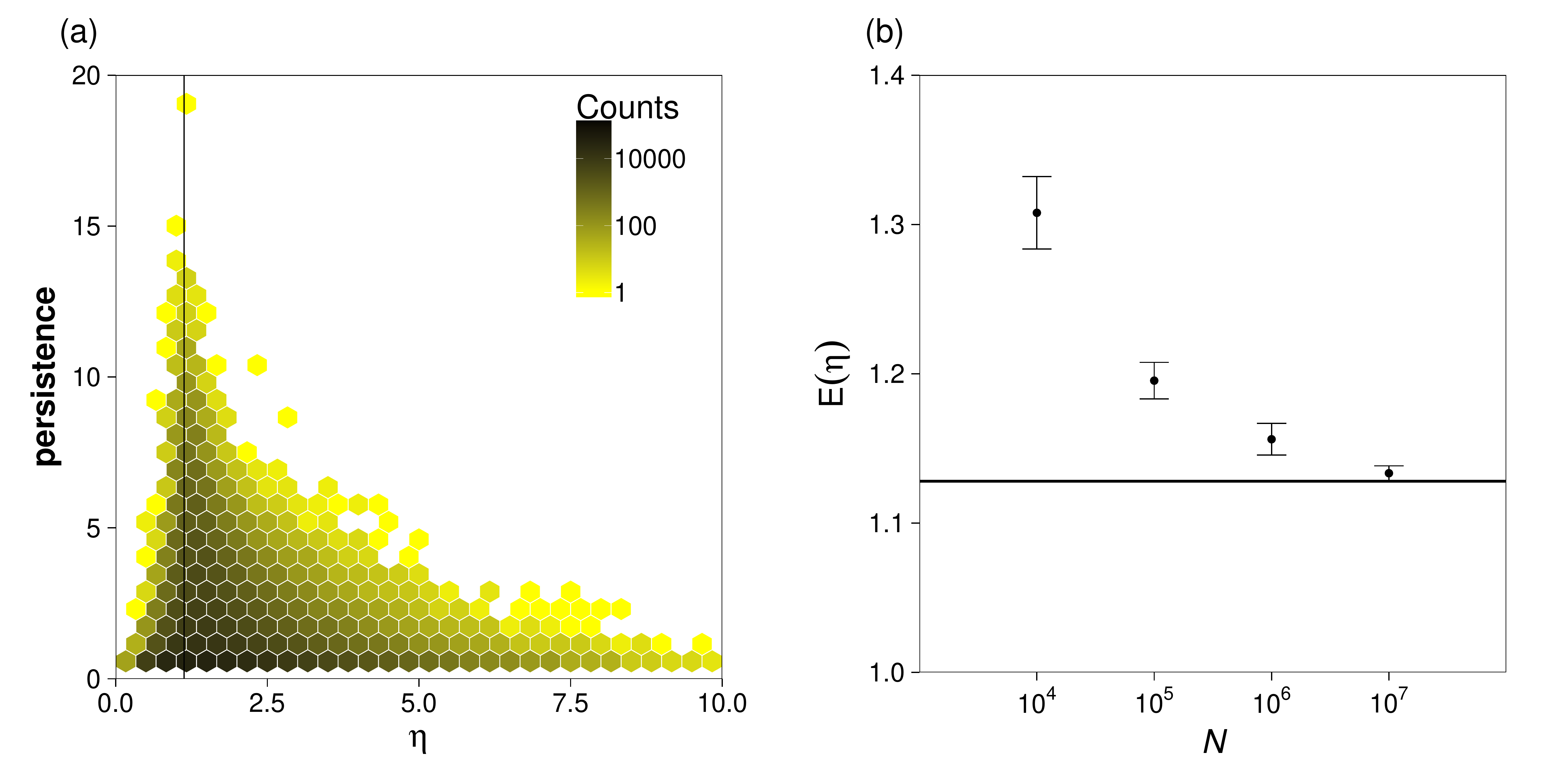}
  \caption{
    \label{fig:birth_persistent_diagram}
    (a) Birth--persistence diagram for 1D invariants of continuum percolation with disks with $N=10^6$ disks, summarized using a hexagonal binning. The vertical black line shows $\eta_{\rm c} = 1.128$. (b) Mean value of $\eta$ of the longest persisting 1D invariant for different values of $N$. The error bars indicate the $95\%$ confidence intervals of the mean. The horizontal line indicates the percolation threshold ($\eta_{\rm c} \approx 1.128$) for $N \to \infty$.
  }
\end{figure*}

\begin{figure*}
  \centering
  \includegraphics[width=\textwidth]{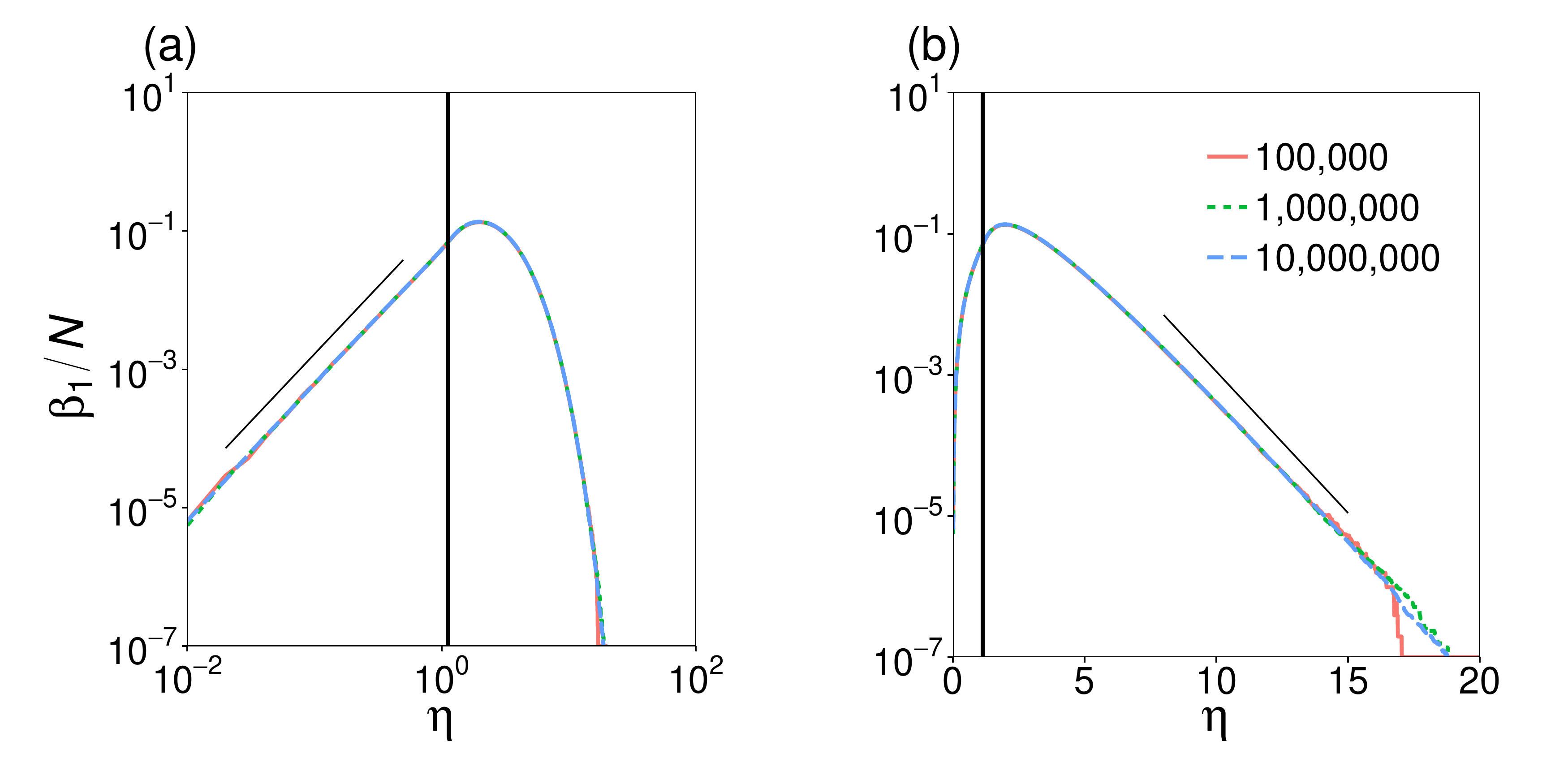}
  \caption{
    \label{fig:betti_number}
    Mean of $\beta_1/N$ as a function of $\eta$ for $N = 10^5$, $N = 10^6,$ and $N = 10^7$ disks in continuum percolation with disks with (a) doubly-logarithmic axes and (b) a logarithmic vertical axis. For reference, we show the percolation threshold $\eta_{\rm c} \approx 1.128$ using a vertical solid line. To guide the eye, we have drawn straight lines in both plots. The slopes of the straight lines are about $1.95$ in panel (a) and about $-0.92$ in panel (b). We calculated these values using a least-squares fit.
  }
\end{figure*}

We aim to apply persistent homology (PH) to characterize structural changes over a wide range of values of $\eta(N,r)$, especially near the percolation transition in the critical regime. PH is a method for topological data analysis (TDA) of point clouds that extracts qualitative features over multiple scales~\cite{Edelsbrunner2010,Zomorodian2005,Otter2015}, which we examine by varying the filling factor $\eta(N,r)$. PH has been applied to investigate force networks in granular material~\cite{Kramar2014,lia2018}, to analyze brain functional networks~\cite{Petri2014}, and for many other applications \cite{Otter2015}.

As we increase the value of $r$, we construct an alpha complex $\mathcal{C}(r)$. By increasing $r$, which causes $\eta(N,r)$ to change, we obtain a nested sequence (a so-called \emph{filtration}) of simplicial complexes. Calculating PH of the filtration then allows us examine how the features evolve with the fill factor $\eta(N,r)$.

Based on the comparison of different implementations of PH by Otter et al.~\cite{Otter2015}, we use the software package Dionysus~\cite{Morozov}. (We test that our results are qualitatively the same when we use the software package (GUDHI)~\cite{Maria2014Gudhi}.) The output from the software consists of pairs of birth and death values. After rescaling, these give, respectively, the values of $\eta(N,r)$ at which a topological feature appears and disappears. 

We visualize the output of PH using a birth--persistence diagram. In this diagram, we plot the persistence of an invariant, defined as the difference of death and birth values of $\eta(N,r)$, versus the value of $\eta(N,r)$ at which the invariant feature is born. In Fig.~\ref{fig:birth_persistent_diagram}(a), we show the birth--persistence diagram of one realization of continuum percolation with disks with $N=10^{6}$ disks. We observe an increasing trend in the persistence of the longest persisting invariants for $\eta(N,r) < \eta_{\rm c}$ and a decreasing trend for $\eta(N,r) > \eta_{\rm c}$. We also see in Fig.~\ref{fig:birth_persistent_diagram}(b) that the longest persisting invariant is born progressively closer to $\eta_{\rm c}$ for larger $N$. This suggests the intriguing possibility that, in the limit $N \to \infty$, the longest persisting invariant is born at $\eta_{\rm c}$. It thereby suggests a possible link between PH in continuum percolation with disks and more conventional approaches to studying such percolation.

In Fig.~\ref{fig:betti_number}, we show how $\beta_1/N$ changes for $\eta(N,r) \in (0,\infty)$. We observe that the curves collapse onto a single curve for a wide range of $\eta(N,r)$, supporting the linear scaling of $\beta_1$ that is predicted by Eq.~\eqref{eq:linear_scaling_beta1}. In Fig.~\ref{fig:betti_number}(a), we see that $\beta_1/N$ increases in a manner that seems to resemble a power law until it peaks at a value of $\eta$ that is larger than the percolation threshold. From Fig.~\ref{fig:betti_number}(b), we observe that it subsequently appears to decay at least exponentially.


\section{Conclusions and Discussion}
\label{sec:discussion}

Continuum percolation with disks is a frequently-studied variant of continuum percolation, and many of its properties have been uncovered using a combination of theoretical and computational techniques~\cite{Meester1996,Gilbert1961,Penrose2003,Kahle2014,Kahle2011}.
In the present paper, we interpreted the union of potentially overlapping disks generated by continuum percolation with disks as a subspace of $[0,1]^2$ and examined its topological properties near the percolation transition. We computed persistent homology and identified structural changes that cannot be described solely by changes in the sizes of clusters. 

We found evidence that the longest persisting invariants are born close to (though seemingly not exactly at) the percolation transition in the limit $N\to \infty$. We also provided numerical confirmation of a theoretical prediction that the first Betti number $\beta_1$ scales linearly with respect to the number $N$ of disks, and we characterized $\beta_1$ as a function of $\eta(N,r)$.

There are several ways to extend our work. It will be interesting, for example, to apply PH to higher-dimensional versions of continuum percolation, and one can also ask how the shapes of objects in percolation models affect topological properties. The latter is reminiscent of important questions in the study of granular materials \cite{lia2018}. Applying PH to related percolation models, such as random sequential adsorption~\cite{Evans1993}, that incorporate correlations between the placement of disks is also likely to be interesting. More generally illustrates that it is useful to examine not only the sizes of components in complex systems, but also their ``shapes'', a perspective that we expect will be insightful for many types of phase transitions.


\section{Acknowledgements}

We thank Bernadette Stolz  and Ulrike Tillmann for helpful comments. LS acknowledges the support provided by the Engineering and Physical Sciences Research Council (EPSRC) through grant number EP/G03706X/1. HAH gratefully acknowledges the EPSRC postdoctoral fellowship EP/KO4196/1 and a Royal Society University Research Fellowship.



%


\end{document}